\documentclass[3p]{elsarticle}

\usepackage{graphicx}
\graphicspath{{figures/}}
\usepackage{caption}
\usepackage{subcaption}
\usepackage{pdfpages}
\usepackage{amsmath}  
\usepackage{nicefrac}
\usepackage{amssymb}          
\usepackage{booktabs}
\usepackage{multirow}
\usepackage{color, colortbl}

\usepackage{blindtext}

\usepackage{lineno,hyperref}
\modulolinenumbers[5]
\hypersetup{pdfauthor=author}

\makeatletter
\newcommand{\setword}[2]{%
  \phantomsection
  #1\def\@currentlabel{\unexpanded{#1}}\label{#2}%
}
\makeatother

\definecolor{DynCyan}{rgb}{0.1058, 0.4549, 0.5686} 
\definecolor{DynRed}{rgb}{0.5529, 0.1215, 0.1333} 
\definecolor{DynGrey}{rgb}{0.3568, 0.4509, 0.5098} 
\definecolor{DynDark}{rgb}{0.0470, 0.2000, 0.2470} 

\journal{Ocean Engineering}

\biboptions{authoryear}

\begin{document}

\begin{frontmatter}


\title{\Large Data assimilation and parameter identification for water waves using the nonlinear Schrödinger equation and physics-informed neural networks}

\author[DYN]{Svenja Ehlers\corref{mycorrespondingauthor}}
\ead{svenja.ehlers@tuhh.de} 
\author[TUD]{Niklas A. Wagner}
\author[TUM]{Annamaria Scherzl}
\author[DLR]{Marco Klein}
\author[DYN,ICL]{Norbert Hoffmann}
\author[TUB]{\mbox{Merten Stender}}

\cortext[mycorrespondingauthor]{Corresponding author}

\address[DYN]{Hamburg University of Technology, Dynamics Group, 21073 Hamburg, Germany}
\address[TUD]{TU Dortmund University, Communication Networks Institute, 44227 Dortmund, Germany}
\address[TUM]{Technical University of Munich, 85748 Garching b. München, Germany}
\address[DLR]{German Aerospace Center, Institute of Maritime Energy Systems, Ship Performance Dep., 21052 Geesthacht, Germany}
\address[ICL]{Imperial College London, Department of Mechanical Engineering, London SW7 2AZ, United Kingdom}
\address[TUB]{Technische Universität Berlin, Cyber-Physical Systems in Mechanical Engineering, 10623 Berlin, Germany}
\begin{abstract}

 The measurement of deep water gravity wave elevations using in-situ devices, such as wave gauges, typically yields spatially sparse data. This sparsity arises from the deployment of a limited number of gauges due to their installation effort and high operational costs. The reconstruction of the spatio-temporal extent of surface elevation poses an ill-posed data assimilation problem, challenging to solve with conventional numerical techniques. To address this issue,  we propose the application of a physics-informed neural network (PINN), aiming to reconstruct physically consistent wave fields between two designated measurement locations several meters apart.
 Our method ensures this physical consistency by integrating residuals of the hydrodynamic nonlinear Schrödinger equation (NLSE) into the PINN's loss function. Using synthetic wave elevation time series from distinct locations within a wave tank, we initially achieve successful reconstruction quality by employing constant, predetermined NLSE coefficients. However, the reconstruction quality is further improved by introducing NLSE coefficients as additional identifiable variables during PINN training. 
The results not only showcase a technically relevant application of the PINN method but also represent a pioneering step towards improving the initialization of deterministic wave prediction methods.

\end{abstract}




\end{frontmatter}

\section{Introduction}

The field of ocean engineering and experimental water wave research highly desires a deterministic description of wave quantities \citep{Klein2020}, which are usually described by partial differential equations (PDEs). Unlike statistical methods, deterministic prediction involves a phase-resolved tracing of wave fields, i.e. the wave elevation $\eta$ as a function of space and time with high resolution. Unfortunately, acquiring such spatio-temporal data from experiments is often impractical: On the one hand, the reconstruction of wave surface elevations from real-world radar data is still an unresolved issue \citep{Ehlers2023}. On the other hand, in-situ measurement devices such as wave gauges measure elevation time-series at a few selected locations only due to the high operational costs. This sparsity of information leads to ill-posed inverse problems \citep{Jagtap2022} when attempting to reconstruct the complete wave elevation $\eta(x,t)$ from gauge measurements $\eta(t)$, which might be solved by numerical or machine learning methods.

Despite substantial advancements in common numerical PDE solvers, such as finite element, finite difference, and spectral methods, their application to highly dynamic systems still incurs remarkable computational costs. These costs primarily stem from the need for fine-grained discretizations to ensure accurate solution approximations. Furthermore, conventional grid-based numerical solvers still face unresolved challenges, especially for addressing ill-posed inverse problems  \citep{Karniadakis2021} such as wave reconstructions. In particular, inverse problems become computationally expensive due to the high number of forward evaluations required to estimate the inverse. Additionally, issues related to numerical stability and convergence necessitate the implementation of suitable regularization techniques \citep{Willard2020}. 

In the meantime, machine learning methods revolutionized many scientific disciplines, including their applications in the field of fast deterministic ocean wave prediction \citep[cf.][]{Mohaghegh2021, Klein2022, Liu2022, Wedler2023} and reconstruction \citep[cf.][]{Ehlers2023, Zhao2023}. However, in contrast to classical numerical solvers, these data-based approaches lack the incorporation of inherent knowledge about the physical laws approximating the underlying system in the form of PDEs. Physical information is solely provided by observational training data for supervised learning, sourced from PDE simulations using classical numerical methods or from field measurements of real physical wave systems. Thus, the solutions generated by data-based approaches may not inherently ensure physical consistency as the quality and quantity of training data limit their accuracy. This can be regarded as a neglect of established knowledge, especially when addressing the reconstruction problem outlined earlier, considering the centuries-long development of model equations for describing surface gravity waves. 

To overcome the limitations of both, explicit numerical PDE solvers and neural network approaches, \cite{Raissi2019} proposed physics-informed neural networks (PINNs). PINNs integrate observational data with physical laws by parameterizing the PDE solution as a neural network to solve forward or inverse problems. More precisely, the network's training process is constrained by a loss function incorporating a PDE residual. To effectively calculate this residual, the PINN algorithm leverages the methods of automatic differentiation \citep{Baydin2018}.
In recent years,  PINNs have gained attention across diverse scientific domains \citep{Karniadakis2021, Cuomo2022}, including computational fluid dynamics \citep{Kissas2020, Raissi2020, Cai2021a}, acoustic wave propagation \citep{RashtBehesht_2022}, heat transfer \citep{Zobeiry2021, Cai2021}, climate modeling \citep{Kashinath2021}, nano-optics \citep{Chen2020} and the study of hyperelastic materials \citep{Nguyen-Thanh2020}. Notably, PINNs have also been employed in a few investigations related to water waves: For instance, \cite{Wang2022} use a loss function based on the wave energy balance equation and the linear dispersion relation to reconstruct near-shore phase-averaged wave heights. The same equations also allow for solving sea bed depth inversion problems from statistical wave parameters in shallow-water regimes \citep{Chen2023}. Furthermore, the Saint-Venant equations within the loss residuals of PINNs enable the downscaling of large-scale river models by assimilating remote sensing data in conjunction with in-situ measurement data of the water surface \citep{Feng2023}. Additionally, the research of \cite{Jagtap2022} showcases the potential of PINNs in resolving ill-posed assimilation problems, leveraging analytical solitary surface measurements and the Serre-Green-Naghdi equations in shallow water scenarios. 

However, in typical ocean engineering research, i.e. down-scaled testing in wave tanks, scenarios arise where the water depth significantly exceeds the wavelength. In such cases, it becomes essential to characterize the nonlinear behavior of water waves in the intermediate to deep water regime \citep{Chiang2005} rather than in shallow water. \cite{Zakharov1968} demonstrated that the amplitude envelope of slowly modulated wave groups approximately satisfies the nonlinear Schrödinger equation (NLSE). Consequently, various variants of the hydrodynamic nonlinear Schrödinger equation have been investigated experimentally and numerically for deterministic wave prediction \citep[cf.][]{Shemer1998, Trulsen2001, Dysthe2003} and rogue wave modeling \citep[cf.][]{Chabchoub2011, Ruban2015}. 

To the best of the authors' knowledge, there is no prior documentation of PINNs to solve the hydrodynamic form of the NLSE, despite the successful integration of other forms of NLSE for different physical phenomena into the PINN framework \cite[cf.][]{Raissi2019, Li2021, Pu2021b, Pu2021a, Wang2021, Jiang2022, Zhang2021}. However, unifying most of this related work demonstrates the application of the NLSE-PINN methodology concerning initial or boundary conditions derived from analytical soliton or breather solutions corresponding to their specific NLSE used in the loss function. While these approaches provide a theoretical proof of concept, they cannot inherently assess the practical viability of PINNs in real-world scenarios, as analytical solutions seldom align with the complexity and imperfections encountered in real measurements.


Therefore, the objective of this study is to demonstrate the application of a NLSE-PINN framework with more realistic data for deep water gravity waves. Despite potential limitations associated with the hydrodynamic NLSE, particularly in accommodating arbitrary irregular sea states due to bandwidth and steepness constraints \citep{Klein2020}, we leverage irregular wave elevation data generated from nonlinear numerical simulations using the high-order spectral method (HOSM) \citep{West1987}. This elevation data is obtained at discrete sampling points, simulating wave gauges within a wave tank, and is considered as authentic measurement data throughout the study. Consequently, we explore the utilization of a potentially slightly misspecified PINN. In this context, the term misspecification refers to the discrepancy between the mathematical model description (e.g. a PDE) and observational data and is a common occurrence for complex physical systems \citep{Zou2023}. Specifically, our research is driven by two hypotheses that are tailored for practical and relevant challenges in ocean engineering and water wave research:

\begin{itemize}
    \item[\setword{H1}{hyp:1}:] PINNs enable the reduction of required wave gauges by assimilating boundary data to reconstruct the fully spatially resolved wave surfaces in between by the physics-based loss function.
    \item[\setword{H2}{hyp:2}:] PINNs facilitate learning optimized NLSE coefficients, thereby enhancing the reconstruction performance compared to constant coefficients manually predetermined from spectral wave properties.
\end{itemize}

To address these hypotheses, we first present the wave tank utilized for wave data generation in Section~\ref{subsec:setup_and_HOS}. Afterwards, the hydrodynamic NLSE is introduced in Section~\ref{subsec:NLSE} to subsequently develop the NLSE-PINN framework and training methodology in Section~\ref{subsec:NLSE-PINN}. The hypotheses are examined in the results Section~\ref{sec:Results}: First, in the assimilation task in Section~\ref{subsec:assimilation}, we evaluate the PINN's capability to reconstruct physically consistent wave surface envelopes from a limited number of gauge measurement points (Hypothesis \ref{hyp:1}). While maintaining constant, predetermined NLSE coefficients in this section, Section~\ref{subsec:coefficient identification} explores the enhancement of reconstruction by incorporating NLSE coefficients as additional trainable PINN variables (Hypothesis \ref{hyp:2}), thereby addressing an inverse problem. Finally, Section~\ref{sec:conclusion} summarizes key findings, outlines method limitations, and suggests potential directions for future research.


\section{Method}
\label{sec:Method}

The following section initiates a concise overview of the wave tank specifications and the numerical method to generate the synthetic wave elevation data. Next, the nonlinear Schrödinger equation for deep water gravity waves is introduced. Subsequently, this equation is integrated to develop a physics-informed neural network, with details on its architecture and training strategy provided in the last subsection.

\subsection{Measurement setup and data generation}
\label{subsec:setup_and_HOS}

\begin{figure}[ht!]
     \centering
     \begin{subfigure}[b]{0.49\textwidth}
         \centering
         \includegraphics[width=\textwidth]{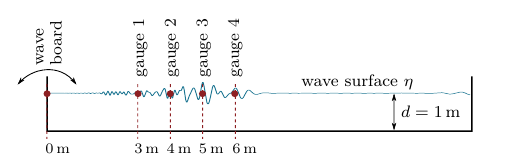}
         \vspace{0.75cm}
         \caption{}
         \label{subfig:wavetank}
     \end{subfigure}
     \begin{subfigure}[b]{0.45\textwidth}
         \centering
         \vspace{-0.40cm}
         \includegraphics[width=\textwidth]{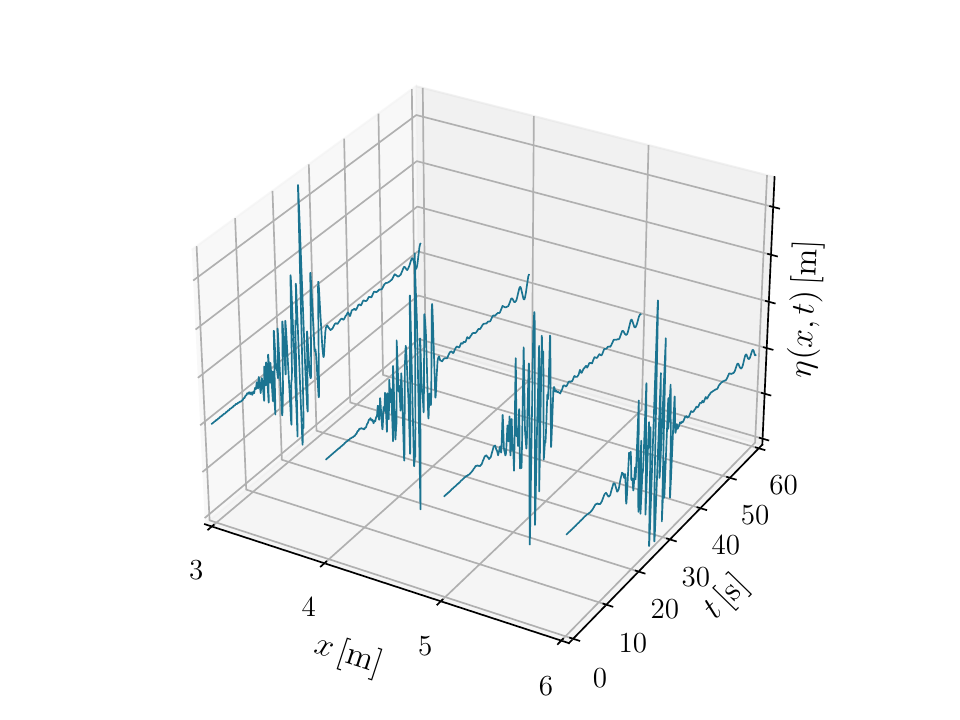}
         \caption{}
         \label{subfig:3D_measurement_visualization}
     \end{subfigure}
     \caption{Measurement setup and corresponding data structure: Wave gauges inside the wave tank acquire the wave elevation at four discrete points in space $x_\mathrm{g}=\{3, 4, 5, 6\} \, \mathrm{m}$ (a), causing a sparse spatio-temporal data structure for each sample (b). }
     \label{fig:SetupMeasurement}
\end{figure}

\begin{figure}[ht!]
\centering
\includegraphics[width=0.45\textwidth]{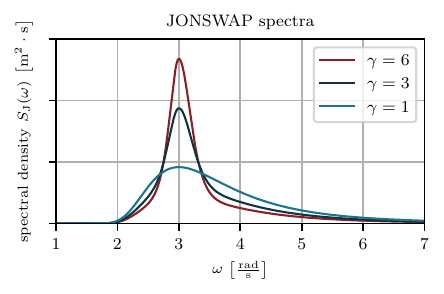}
\vspace{-0.2cm}
     \caption{JONSWAP spectra for different peak enhancement factors $\gamma$ used to initialize the HOSM wave simulations. Higher values of $\gamma$ result in narrower spectra for the generated irregular waves.}
     \label{fig:JONSWAP}
\end{figure}

The numerical wave tank experiments are based on a wave tank facility at Hamburg University of Technology. This wave tank possesses a cross-sectional area of $1.5 \, \mathrm{m} \times 1.5 \, \mathrm{m}$ and extends over a length of $15 \, \mathrm{m}$ \citep{Klein2022}. To generate waves, a flap-type board is installed on one side of the tank,  while the opposite side has a beach element to minimize wave reflections. In our investigations we maintain a fixed water depth of $d=1 \, \mathrm{m}$, and positioned four wave gauges at specific locations $x_\mathrm{g} \in \{3, 4, 5, 6\} \, \mathrm{m}$ on the water's surface. At the gauge locations we simulate temporal series of wave surface elevations $\eta(x=x_\mathrm{g}, t)$ over a temporal interval spanning $t= 0-60 \, \mathrm{s}$. A graphical representation of this measurement setup can be found in Fig.~\ref{subfig:wavetank}, while the resulting data exhibits a three-dimensional structure, as exemplified in Fig.~\ref{subfig:3D_measurement_visualization}.
The irregular one-dimensional nonlinear water wave surfaces are modeled on a Cartesian coordinate system $(x,z)$ assuming a Newtonian fluid that is incompressible, inviscid, and irrotational, by the general initial boundary-value potential flow problem modeled by the Laplace equation
\begin{align}
    \nabla^2 \Phi = \Phi_{xx} + \Phi_{zz}&=0  \label{eq:Laplace}
\end{align}
and the boundary conditions
\begin{align}
    \eta_t +  \eta_x \Phi_x - \Phi_z&=0 \hspace{0.5cm}      \text{on } z=\eta(x, t)  \label{eq:kinBC}\\ 
    \Phi_t +g \eta   + \frac{1}{2} \left(\Phi_{xx}^2 +\Phi_{zz}^2  \right)&=0 \hspace{0.5cm} \text{on } z=\eta(x, t) \label{eq:dynBC}\\
    \Phi_z &=0  \hspace{0.5cm}  \text{on } z=-d \label{eq:bot} 
\end{align}
within the fluid domain. Therein, $\Phi (x,z,t)$ is the velocity potential, $\eta(x,t)$ is the free surface elevation, and $g$ is the gravity acceleration. The mean free surface is located at $z=0 \, \mathrm{m}$ with $z$ pointing upwards. We employ the high-order spectral method (HOSM) of order $M=4$ as formulated by \cite{West1987} to simulate nonlinear wave propagation. The simulations are initialized using wave surfaces acquired from JONSWAP spectra \citep{Hasselmann1973} in finite depth form \citep{Bouws1985}. To generate wave data covering different wave conditions and significant wave heights $H_\mathrm{s}$, the sea state parameters of peak wave frequency $\omega_\mathrm{p}= \nicefrac{2 \pi}{T_\mathrm{p}}$ (with $T_\mathrm{p}$ being the peak period), wave steepness $\epsilon=k_\mathrm{p} \nicefrac{H_\mathrm{s}}{2}$ (with $k_\mathrm{p}= \nicefrac{2 \pi}{L_\mathrm{p}}$ being the peak wavenumber and $L_\mathrm{p}$ being the peak wavelength) and peak enhancement factor $\gamma$ are varied across 
\begin{align*}
\omega_\mathrm{p} &\in \{3, \, 4, \, 5, \, 6, \, 7, \, 8, \, 9\} \, \tfrac{\mathrm{rad}}{\mathrm{s}} \\
\epsilon & \in \{0.0125, \, 0.0250, \, 0.0375, \, 0.0500, \, 0.0750, \, 0.1000 \}\\
\gamma & \in \{1, \, 3, \,6 \},
\end{align*}
whereas higher $\gamma$ indicate narrow-banded spectra as shown in Fig.~\ref{fig:JONSWAP} . By randomly selecting initial phase shifts of the component waves, five different elevations are generated for each $\epsilon$-$\omega_\mathrm{p}$-$\gamma$-combination, resulting in 630 different wave samples in total. For details on the nonlinear wave data generation process in a wave tank facility using the HOSM, the reader is referred to \cite{Klein2022} and \cite{Luenser2022}.

\subsection{Hydrodynamic nonlinear Schrödinger equation}
\label{subsec:NLSE}

In addition to employing numerical methods such as HOSM to approximate the potential flow equations~\eqref{eq:Laplace}-\eqref{eq:bot}, an alternative approach is the derivation of simplified solutions. By utilization of perturbation theory around the parameters of wave steepness $\epsilon=k_\mathrm{p} a$ and relative bandwidth $\mu = \nicefrac{\Delta k}{k_\mathrm{p}} <<1$, small amplitude waves and narrow bandwidth are assumed. Moreover, the boundary value problem at the unknown free surface $z=\eta(x,t)$ can be approximated using a Taylor series expansion. By truncating the perturbation expansion at order $\mathcal{O}(\epsilon^3)$, the envelope equation of the nonlinear Schrödinger equation (NLSE) is derived \citep{Zakharov1968, Hasimoto1972, Yuen1982} in terms of a complex wave envelope amplitude $A(x,t)= U(x,t) + i V(x,t)$ that varies slowly compared to the phase $\vartheta = k_\mathrm{p} x - \omega_\mathrm{p} t $ of its underlying carrier wave $\eta(x,t)$. The hydrodynamic NLSE in time-like form
\begin{equation}
    i\left( A_x +\frac{1}{c_\mathrm{g}} A_t \right) + \delta A_{tt} +\nu |A|^2 A = 0 \label{eq:NLSE}
\end{equation}
finds common application in boundary-value wave tank problems \citep{Chabchoub2016}, whereas
\begin{equation}
    c_\mathrm{g}= \frac{\omega_\mathrm{p}}{2 k_\mathrm{p}}, \:\:\: \delta = - \frac{k_\mathrm{p}}{\omega_\mathrm{p}^2}, \:\:\: \nu = - k_\mathrm{p}^3,
    \label{eq:NLSE_coeeffs}
\end{equation} 
are the NLSE coefficients. The peak frequency $\omega_\mathrm{p}$ of the carrier wave is related to the corresponding peak wavenumber $k_\mathrm{p}$ through the linear dispersion relation 
\begin{equation}
\omega_\mathrm{p} = \sqrt{g k_\mathrm{p} }.
\label{eq:linear_dispersion}
\end{equation}
The first term in Eq. \eqref{eq:NLSE} characterizes the spatial variation of the amplitude, while the second term represents the wave propagation by the group velocity $c_\mathrm{g}$. The third term introduces the dispersive effect, while the last term introduces the nonlinearity. 
In contrast to the potential flow equations, which require an approximation in the depth direction, the NLSE is regarded as the simplest nonlinear equation for deep-water wave dynamics, as it is uniquely formulated on the water surface \citep{Osborne2010}.
However, while capturing essential aspects of nonlinear water waves, the NLSE's limitations in nonlinearity magnitude and spectral bandwidth limit its general applicability in deterministic wave prediction \citep{Klein2020}.\\

\begin{figure}[ht!]
\centering
\includegraphics{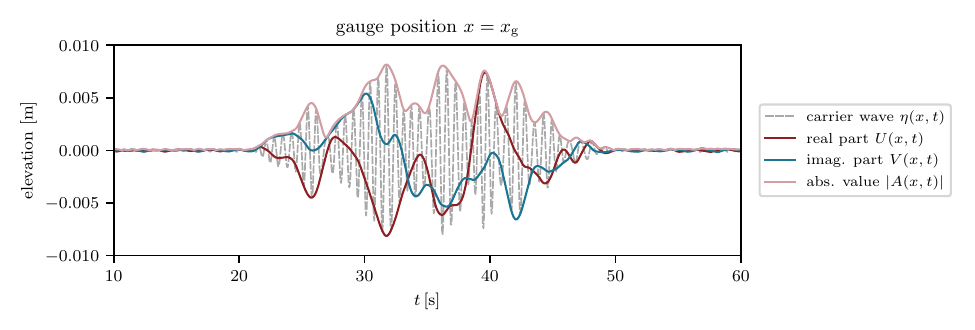}
\vspace{-0.75cm}
     \caption{Example of an irregular carrier wave $\eta(x,t)$ 
      measured at one location inside the wave tank. The real and imaginary part of its corresponding complex envelope $A(x,t) = U(x,t) + i V(x,t)$ are visualized, along with its absolute value $|A(x,t)|$.}
     \label{fig:carrier_envelope}
\end{figure}

In practical cases, where real-valued measured time series $\eta(x=x_\mathrm{g}, t)$ are available, the values of $\omega_\mathrm{p}$ and $k_\mathrm{p}$ are commonly derived from a spectral representation via discrete Fourier transform $F(\omega)$ and the dispersion relation in Eq. \eqref{eq:linear_dispersion}. We employ an advanced method based on \cite{Sobey1986}, defining
\begin{equation}
\omega_\mathrm{p} =\frac{\int \omega \cdot F(\omega)^5 d \omega}{\int F(\omega)^5 d \omega},
\label{eq:omega_ref}
\end{equation}
which enhances robustness compared to determining $\omega_\mathrm{p}$ as the frequency where $F(\omega)$ attains its maximum.
The complex amplitudes for initializing the NLSE are computed as
\begin{equation}
    A(x_\mathrm{g}, t) = \left[ \eta(x_\mathrm{g},t) +i \mathcal{H} \left( \eta(x_\mathrm{g},t) \right) \right] \cdot \exp(-i \vartheta), \label{eq:Hilbert-Transform}
\end{equation}
where $\mathcal{H}$ denotes the Hilbert transform of the measured signal \citep{Huang1999, Thrane2011}. The other way around, the equation
\begin{equation}
\eta(x_\mathrm{g},t)=\mathrm{Re}\left[A(x_\mathrm{g},t) \cdot \exp(i \vartheta) \right]  
\label{eq:backtransform}
\end{equation}
instead allows for the reverse transformation from a complex envelope back to a carrier wave elevation after the reconstruction procedure, although this transformation is not presented in this work. The relationship between carrier wave $\eta(x_\mathrm{g},t)$ and complex envelope $A(x_\mathrm{g},t)=U(x_\mathrm{g},t) + i V(x_\mathrm{g},t)$ is illustrated in Fig.~\ref{fig:carrier_envelope}.

\subsection{Physics-informed neural network for the NLSE}
\label{subsec:NLSE-PINN}

As discussed in the previous section, the hydrodynamic NLSE exhibits inherent simplifications concerning the generalized water wave problem. Consequently, we undertake an investigation to assess the degree to which this equation can effectively serve within a PINN for the analysis of surface waves arising from different wave physics. This is motivated by the NLSE's simple definition, limited to the $(x, t)$-domain only. Therefore, the following sections introduce the architecture, loss function, training methodology, and evaluation metrics adopted for our NLSE-PINN framework.

\subsubsection{PINN architecture}
The neural network architecture of the PINN developed in our study is depicted on the lower left side of Fig.~\ref{fig:PINN}.  It includes two input nodes for collocation points in space $x$ and time $t$. Given the complex-valued solution $A(x,t) = U(x,t) + i V(x,t)$ of the NLSE, two output nodes are required \citep{Raissi2019}, where $U$ represents the real part and $V$ the imaginary part. Moreover, the network consists of four intermediate hidden layers, each comprising 200 nodes, resulting in a total network depth of $D=6$. The variables of each layer $l$, denoted as $\boldsymbol{\theta}^{(l)} = \{\mathbf{W}^{(l)}, \mathbf{b}^{(l)} \}_{1\le l \le D}$, receive their initial values using the Xavier initialization technique for weights $\mathbf{W}^{(l)}$ \citep{Glorot2010}, while biases $\mathbf{b}^{(l)}$ are initialized to zero. The total amount of the network's weights and biases is denoted as $\boldsymbol{\theta}= \left[\boldsymbol{\theta}^{(1)}, \hdots ,\boldsymbol{\theta}^{(D)} \right]$.

To enhance convergence and accuracy during the training, we incorporate a strategy using layer-wise locally adaptive activation functions \citep{Jagtab2020b, Jagtap2020c}, which has shown promising applications i.e. in \cite{Pu2021a}, \cite{Jagtap2020a}, \cite{Jagtap2022}, \cite{Shukla2020} and \cite{Guo2023}. The input of a hidden layer $\mathbf{h}^{(l+1)}$ is derived from the output of the preceding layer $\mathbf{o}^{(l)}$ following the rule
\begin{align}
\mathbf{o}^{(l)} &= \mathbf{W}^{(l)\mathrm{T}} \cdot \mathbf{h}^{(l)} +\mathbf{b}^{(l)}\\
  \mathbf{h}^{(l+1)} &= \tanh{(s \cdot a^{(l)} \cdot \mathbf{o}^{(l)}) },
\end{align}
for $l=1, \hdots, D-1$. The last layer $D$ has a linear activation function. In this context, $\mathbf{W}^{(l)}$ and $\mathbf{b}^{(l)}$ are the weight and bias matrices of the $(l)$-th layer, and $s=10$ is a fixed scaling factor. Furthermore, $\boldsymbol{a}=~\left[a^{(1)}, \hdots, a^{(D-1)} \right]$ represent additional trainable variables of the network, which are fine-tuned during the training process to modulate the slopes of the activation functions. In our case, we initialize with $a^{(l)}=0.2$, resulting in the initial slope of $s \cdot a^{(l)}=2$ being slightly steeper than the usual $\tanh$ activation.

\subsubsection{PINN loss function and training}
\begin{figure}[ht]
\centering
\includegraphics{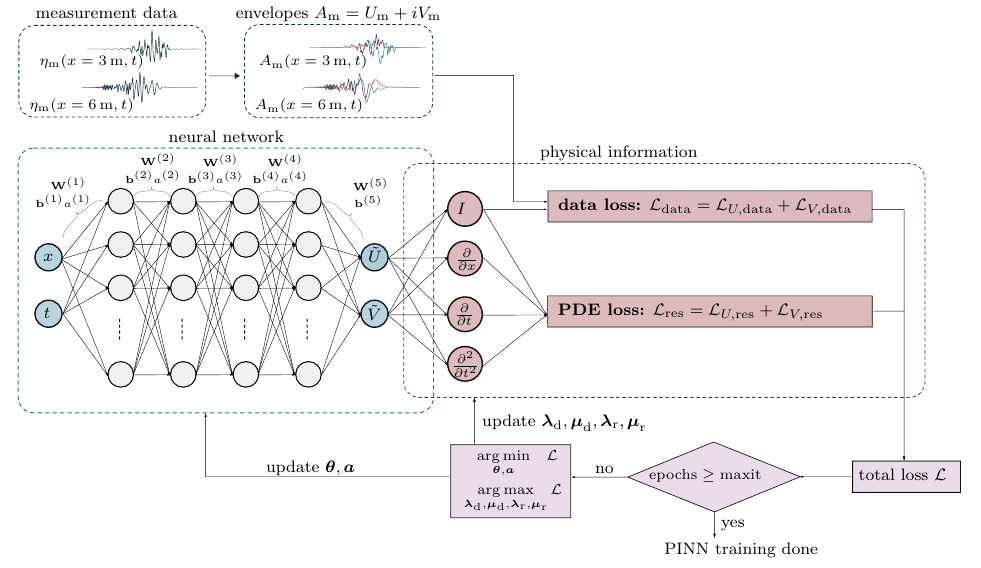}
\caption{Schematic framework of the physics-informed neural network developed to solve the hydrodynamic nonlinear Schrödinger equation. The neural network architecture comprises two input nodes to insert points of the computational domain $(x,t)$ and two output nodes to approximate the real- and imaginary part of the complex-valued NLSE solution $\Tilde{A}(x,t)=\Tilde{U}(x,t)+i\Tilde{V}(x,t)$. Real wave measurement data $\eta_\mathrm{m}$, obtained at the domain boundaries, is transformed into envelope representations $A_\mathrm{m}$ to guide the PINN's solution towards approximating these boundary values. This is achieved by the data loss component $\mathrm{MSE}_\mathrm{data}$ (Eq. \eqref{eq:MSE_Udata}- \eqref{eq:MSE_Vdata}). To additionally guide the PINN solution towards ensuring physical consistency inside the entire computational domain, the PDE-loss $\mathrm{MSE}_\mathrm{res}$ incorporates NLSE residuals (Eq. \eqref{eq:MSE_Ures}-\eqref{eq:res_V}). The PINNs variables $\boldsymbol{\theta}$ (weights $\mathbf{W}$ and biases $\mathbf{b}$), activation function slopes $\boldsymbol{a}$ and self-adaption weights $\boldsymbol{\lambda}_{\mathrm{d}}, \boldsymbol{\mu}_{\mathrm{d}}, \boldsymbol{\lambda}_{\mathrm{r}}, \boldsymbol{\mu}_{\mathrm{r}}$ are updated iteratively in the training process to minimize the total loss $\mathcal{L}$, which is composed of the data and PDE loss component.}
\label{fig:PINN}
\end{figure}

To enable the PINN's capability to approximate a NLSE surrogate model $\tilde{A}(x,t)= \tilde{U}(x,t) + i\tilde{V}(x,t)$ that is parameterized by $\boldsymbol{\theta}$ and $\boldsymbol{a}$, we use wave elevation measurement data $\eta_\mathrm{m} \in \mathbb{R}^{N_\mathrm{d}}$ from two outer gauge positions, $x_\mathrm{g,meas} \in \{3, 6\} \, \mathrm{m}$, within the tank. The elevation data at the remaining locations, $x_\mathrm{g,test} \in \{4, 5\} \, \mathrm{m}$, is reserved for later evaluation and not incorporated in the training process. Each elevation measurement is transferred to a complex envelope $A_\mathrm{m} \in \mathbb{C}^{N_\mathrm{d}}$ using Eq.~\eqref{eq:Hilbert-Transform}, as depicted in the top boxes of Fig.~\ref{fig:PINN}. In this work,  $N_\mathrm{d}=1,200$ is the total number of measurement data points at the domain boundaries, denoted as  $\{x_\mathrm{d}= x_\mathrm{g,meas}, t_\mathrm{d}^{(j)} \}_{j=1}^{N_\mathrm{d}}$. These points are obtained by sampling the temporal sequence $t=0-60 \, \mathrm{s}$ with an increment of $\Delta t = 0.05 \, \mathrm{s}$. In addition, we incorporate a set of $N_\mathrm{r}=20,000$ randomly located collocation points $\{x_\mathrm{r}^{(j)}, t_\mathrm{r}^{(j)} \}_{j=1}^{N_\mathrm{r}}$ to enforce the NLSE solution across the entire computational $(x,t)$-domain. Using these measurements and sets of points, the multi-objective PINN-loss function is formally defined by 
\begin{align}
\mathcal{L} &= \mathcal{L}_{U,\mathrm{data}} + \mathcal{L}_{V,\mathrm{data}} + \mathcal{L}_{U,\mathrm{res}} + \mathcal{L}_{V,\mathrm{res}}
\end{align}
whereas the loss components are
\begin{align}
  \mathcal{L}_{U,\mathrm{data}} &=  \underbrace{\frac{1}{N_\mathrm{d}} \sum_{j=0}^{N_\mathrm{d}} \left|\tilde{U}(x_\mathrm{d}, t_\mathrm{d}^{(j)})- U_\mathrm{m}^{(j)}\right|^2}_{\mathrm{MSE}_{U,\mathrm{data}}} \cdot \left(\lambda_{\mathrm{d}}^{(j)} \right)^2\label{eq:MSE_Udata}\\
\mathcal{L}_{V,\mathrm{data}} &= \underbrace{ \frac{1}{N_\mathrm{d}} \sum_{j=0}^{N_\mathrm{d}} \left|\tilde{V}(x_\mathrm{d}, t_\mathrm{d}^{(j)})- V_\mathrm{m}^{(j)}\right|^2}_{\mathrm{MSE}_{V,\mathrm{data}}} \cdot \left(\mu_{\mathrm{d}}^{(j)} \right)^2\label{eq:MSE_Vdata}\\
\mathcal{L}_{U,\mathrm{res}} &= \underbrace{\frac{1}{N_\mathrm{r}} \sum_{j=0}^{N_\mathrm{r}} \left|\tilde{R_U}(x_\mathrm{r}^{(j)}, t_\mathrm{r}^{(j)})\right|^2}_{\mathrm{MSE}_{U,\mathrm{res}}} \cdot  \left(\lambda_{\mathrm{r}}^{(j)} \right)^2 \label{eq:MSE_Ures}\\
 \mathcal{L}_{V,\mathrm{res}} &= \underbrace{\frac{1}{N_\mathrm{r}} \sum_{j=0}^{N_\mathrm{r}} \left|\tilde{R}_V(x_\mathrm{r}^{(j)}, t_\mathrm{r}^{(j)})\right|^2}_{ \mathrm{MSE}_{V,\mathrm{res}}} \cdot \left(\mu_{\mathrm{r}}^{(j)} \right)^2. \label{eq:MSE_Vres}
\end{align}

The terms $\mathrm{MSE}_{U,\mathrm{data}}$ and $\mathrm{MSE}_{V,\mathrm{data}}$ serve to quantify the error between PINN predictions $(\Tilde{U}, \Tilde{V})$ and measured envelope data $(U_\mathrm{m}, V_\mathrm{m})$ at the domain boundaries. On the other hand, $\mathrm{MSE}_{U,\mathrm{res}}$ and $\mathrm{MSE}_{V,\mathrm{res}}$ describe the extent to which the PINN solution aligns with the NLSE residuals
\begin{align}
 \tilde{R}_U(x,t):=&-\tilde{V}_x(x,t) -\frac{1}{c_g} \tilde{V}_t(x,t) + \delta \tilde{U}_{tt}(x,t) +\nu \left(\tilde{U}(x,t)^2+ \tilde{V}(x,t)^2\right) \cdot \tilde{U}(x,t). \\
 \tilde{R}_V(x,t):=& \hspace{0.4cm} \tilde{U}_x(x,t) +\frac{1}{c_g} \tilde{U}_t(x,t) + \delta \tilde{V}_{tt}(x,t) +\nu \left(\tilde{U}(x,t)^2+ \tilde{V}(x,t)^2\right) \cdot \tilde{V}(x,t) \label{eq:res_V}
\end{align}
These residuals are calculated using automatic differentiation \citep{Baydin2018} of the neural network's output with respect to the input variables $x$ and $t$, as illustrated on the right of Fig. \ref{fig:PINN}. Furthermore, following the approach proposed by \cite{McClenny2020}, trainable self-adaptation weights $\lambda_\mathrm{d}^{(j)}$, $\mu_\mathrm{d}^{(j)}$, $\lambda_\mathrm{r}^{(j)}$ and $\mu_\mathrm{r}^{(j)}$ are introduced in the loss components (Eqs.\eqref{eq:MSE_Udata} -\eqref{eq:MSE_Vres}). These variables, initially set to one, are associated with each measurement point $\{x_\mathrm{d}, t_\mathrm{d}^{(j)} \}_{j=1}^{N_\mathrm{d}}$ or collocation point $\{x_\mathrm{r}^{(j)}, t_\mathrm{r}^{(j)} \}_{j=1}^{N_\mathrm{r}}$. The PINN autonomously identifies challenging regions inside the solution domain characterized by high point-specific errors and increases the respective self-adaptation weights to emphasize the penalty and thus improve the approximation. The behavior is achieved through the concurrent minimization of the loss function $\mathcal{L} =\mathcal{L}(\boldsymbol{\theta}, \boldsymbol{a},\boldsymbol{\lambda}_{\mathrm{d}}, \boldsymbol{\mu}_{\mathrm{d}}, \boldsymbol{\lambda}_{\mathrm{r}}, \boldsymbol{\mu}_{\mathrm{r}})$ by updating the network weights and biases $\boldsymbol{\theta}$ and activation slopes $\boldsymbol{a}$ alongside with the maximization of loss for the self-adaption weights $\boldsymbol{\lambda}_\mathrm{d}, \boldsymbol{\mu}_\mathrm{d}, \boldsymbol{\lambda}_\mathrm{r}, \boldsymbol{\mu}_\mathrm{r}$ in each epoch of training.

In many instances, PINNs are trained using a two-step strategy involving the Adam optimizer \citep{Kingma2014} for a defined number of epochs, followed by the L-BFGS optimizer \citep{Liu1989}.  This strategy is recognized as pivotal \citep{Markidis2021}, with Adam preventing convergence into local minima initially, while L-BFGS refines small-scale solution components \citep{Cuomo2022}. However, this approach, beneficial for smooth analytical solutions, proved less suitable for our studies: The L-BFGS optimizer tends to over-refine noisy elements within measurements. We found that the AMSGrad modification of the Adam optimizer \citep{Reddi2019} from the PyTorch library \citep{Pytorch2019} yields satisfactory results with a learning rate of $\alpha = 0.0001 $ over $15,000$ epochs of training.

\subsubsection{Evaluation}
\label{sec:evaluation}
After assimilating the measurements at the domain boundaries to reconstruct the envelopes in the entire computational domain, the evaluation of the PINNs' performance necessitates a metric that compares the measurement time-series $A_\mathrm{m}=A_\mathrm{m}(x=x_\mathrm{g},t)$ with the reconstruction $\Tilde{A}=\Tilde{A}(x=x_\mathrm{g},t)$ at all gauge positions $x_\mathrm{g} \in  \{3, 4, 5, 6 \} \, \mathrm{m}$. While metrics based on Euclidean distances are scale-depended and treat deviations in frequency and phase as amplitude errors \citep{Wedler2022}, the surface similarity parameter (SSP)
\begin{equation}
    \mathrm{SSP}(A_\mathrm{m}, \Tilde{A}) = \frac{\sqrt{\int | F_{A_\mathrm{m}}(\omega)-F_{\Tilde{A}}(\omega)|^2 d \omega}}{\sqrt{\int | F_{A_\mathrm{m}}(\omega)|^2 d \omega}+\sqrt{\int |F_{\Tilde{A}}(\omega)|^2 d \omega}} \in [0, 1],
    \label{eq:SSP}
\end{equation}
 proposed by \cite{Perlin2014} combines phase-, amplitude-, and frequency errors into a scalar unified measure. In this metric, $\omega$ denotes the wave frequency vector and $F_{A_\mathrm{m}}$ and $F_{\Tilde{A}}$ denote the discrete Fourier transforms of the time-series $A_\mathrm{m}(x=x_\mathrm{g},t)$ or $\Tilde{A}(x=x_\mathrm{g},t)$. The SSP is a normalized error metric, with $\mathrm{SSP}=0$ indicating perfect agreement and $\mathrm{SSP}=1$ implying a comparison against zero or phase-inverted surfaces. Due to this straightforward error assessment, the SSP has found application in recent research related to ocean wave prediction and reconstruction by \cite{Klein2020, Klein2022}, \cite{Wedler2022, Wedler2023}, \cite{Desmars2021, Desmars2022}, \cite{Luenser2022}, \cite{Kim2023} and \cite{Ehlers2023}. Empirical investigations revealed that $\mathrm{SSP} \le 0.2$ can be considered satisfactory within this study.


\section{Results and discussion}
\label{sec:Results}

In the following section, the NLSE-PINN framework undergoes training for all 630 generated wave measurement samples, each characterized by a $\epsilon$-$\omega_\mathrm{p}$-$\gamma$-combination. The results are presented systematically and analyzed in the context of NLSE physics and the underlying data generated by the HOSM. The first subsection focuses on a pure data assimilation task, where NLSE coefficients are kept constant and determined a-priori based on spectral wave properties and linear dispersion. This facilitates the reconstruction of envelopes across the entire spatial domain between two time-series measured at locations that are $3 \, \mathrm{m}$ apart. In contrast, in the second subsection, we treat NLSE coefficients as trainable variables, constituting an additional coefficients identification task while assimilating measurement data. This approach serves to reduce uncertainties associated with the coefficients, leading to improved reconstruction results.

\subsection{Data assimilation}
\label{subsec:assimilation}

As posited in Hypothesis~\ref{hyp:1}, the objective of the data assimilation task is to reduce the number of costly wave elevation sensors by replacing them with surface reconstructions generated by the NLSE-PINN. To achieve this, gauges for measuring $\eta_\mathrm{m}$ are limited to two specific locations $x_\mathrm{g,meas} \in \{3, 6\} \, \mathrm{m}$ in our experimental setup in Fig.~\ref{fig:SetupMeasurement}. This requires the approximation of the surface elevation within the entire computational domain $3 \, \mathrm{m} \le x \le 6 \, \mathrm{m}$ and $0 \, \mathrm{s}\le t \le 60 \, \mathrm{s}$ solely based on boundary data. Importantly, additional measurements at locations $x_\mathrm{g, test} \in \{4, 5\} \, \mathrm{m}$ are reserved exclusively to evaluate PINN solutions in terms of SSP values and are not integrated into the training process.

For each of the 630 samples, the NLSE coefficients (Eq. \eqref{eq:NLSE_coeeffs}) are determined a-priori based on the mean of the spectral peak frequencies $\omega_{\mathrm{p},3}$ and $\omega_{\mathrm{p},6}$ from the measured elevation time series at the domain boundaries using Eq. \eqref{eq:omega_ref}. This peak frequency $\overline{\omega}_\mathrm{p} = \tfrac{1}{2} \left(\omega_{\mathrm{p},3}+\omega_{\mathrm{p},6}\right)$ might deviate from the peak frequency $\omega_\mathrm{p}$ used to generate the JONSWAP spectrum for the HOSM. The Eq. \eqref{eq:linear_dispersion} yield the corresponding peak wavenumber $\overline{k}_\mathrm{p}$. As the NLSE is defined in terms of a complex envelope $A_\mathrm{m}$, all representations of carrier wave elevations $\eta_\mathrm{m}$ are transformed using Eq. \eqref{eq:Hilbert-Transform}. Each sample undergoes an individual PINN training for 15,000 epochs, requiring approximately $1,200 \, \mathrm{s}$ of computational time on a NVIDIA GeForce RTX 3090 GPU. \\

\begin{figure}[ht!]
\centering
\includegraphics[width=0.7\textwidth]{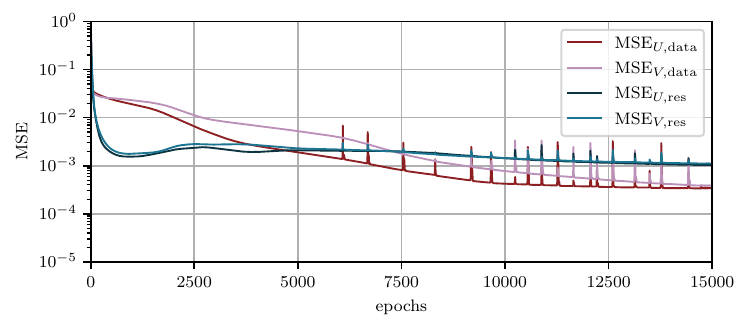}
\vspace{-0.3cm}
     \caption{Exemplary NLSE-PINN training loss curve. The PDE residual error components $\mathrm{MSE}_{U,\mathrm{res}}$ and $\mathrm{MSE}_{V,\mathrm{res}}$ initially exhibit a strong decrease, but slightly increase as the data error components $\mathrm{MSE}_{U,\mathrm{data}}$ and $\mathrm{MSE}_{V,\mathrm{data}}$ are reduced. After around 10,000 epochs, the PDE residual errors reach a plateau, while the data errors continue to gradually improve.}
     \label{fig:example_loss}
\end{figure}

Fig.~\ref{fig:example_loss} depicts a representative training loss curve for the NLSE-PINN framework. During the initial epochs, a remarkable decrease in the PDE residual error components $\mathrm{MSE}_{U,\mathrm{res}}$ and $\mathrm{MSE}_{V,\mathrm{res}}$ is observed, while data errors $\mathrm{MSE}_{U,\mathrm{data}}$ and $\mathrm{MSE}_{V,\mathrm{data}}$ remain high. As the data errors start decreasing and the PINN solution aligns more closely with the prescribed boundary conditions, the PDE errors experience a transient increase before reaching a plateau. These observations may arise from a zero envelope causing high data errors in the early training, despite satisfying PDE residues causing low PDE errors. 
However, as the envelope progressively matches the actual measurement data at the boundaries, fulfilling PDE residues in the remaining domain becomes more challenging, leading to a slight increase in the corresponding error. 

\begin{figure}[ht!]
     \centering
     \vspace{0.2cm}
     \includegraphics[width=0.97\textwidth]{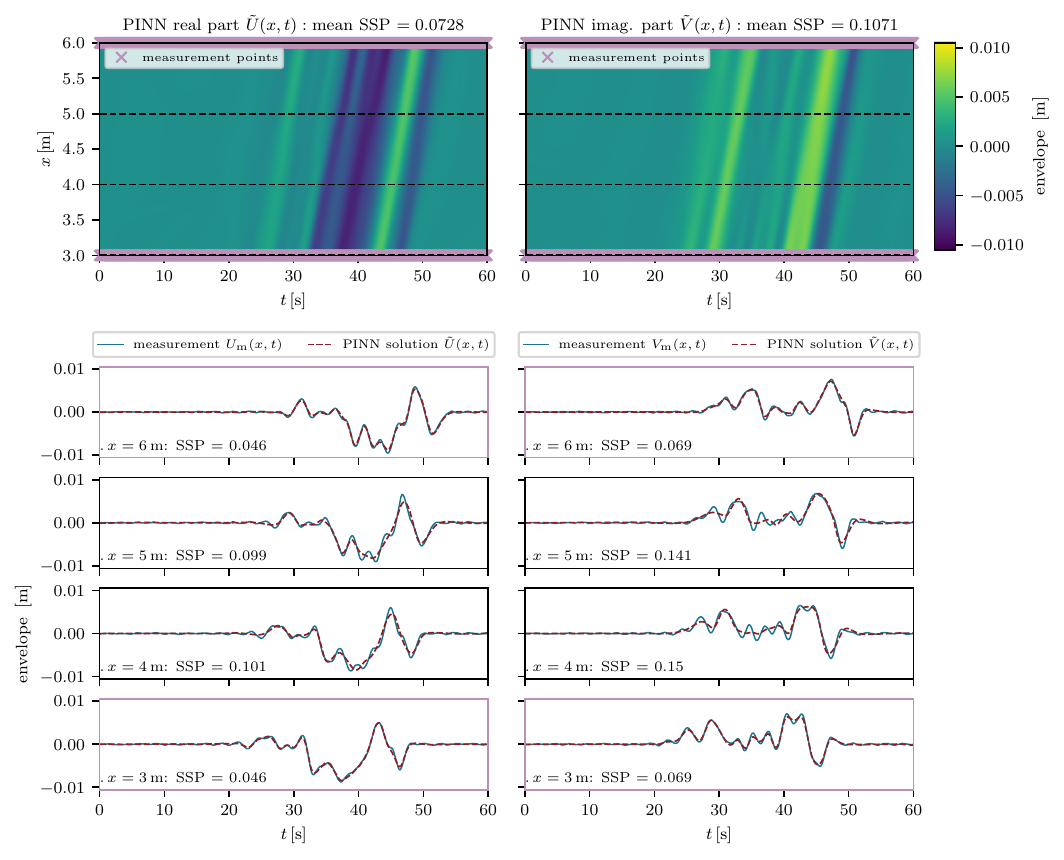}
     \caption{Real part (a) and imaginary part (b) of the envelope solution $\Tilde{A}(x,t) = \Tilde{U}(x,t) + i \Tilde{V}(x,t)$ generated by the PINN using predetermined, constant NLSE coefficients for sample no. 1 (carrier wave with $\epsilon=0.075$, $\overline{\omega}_\mathrm{p}= 9.45 \, \frac{\mathrm{rad}}{\mathrm{s}}$, $\gamma=6$). Measurement envelope points $U_\mathrm{m}, V_\mathrm{m}$ were provided on the domain boundaries only, highlighted in violet. While the PINN solution at the boundaries aligns well with the measured data, it exhibits slight inaccuracies within the remaining computational domain.}
     \label{fig:samp1}
     \vspace{-0.2cm}
\end{figure}

In Fig.~\ref{fig:samp1}, we present a reconstruction of the real part $\Tilde{U}(x,t)$ and imaginary part $\Tilde{V}(x,t)$ of an envelope above the carrier wave sample no. 1 with $\epsilon=0.075$, $\overline{\omega}_\mathrm{p}= 4.45 \, \frac{\mathrm{rad}}{\mathrm{s}}$ and $\gamma=6$. The PINN reconstructs a wave envelope structure across the entire computational domain, despite relying solely on measurement data from the boundaries marked in violet.
The consistency of the reconstructed envelope is evaluated by comparing the PINN's solution $\Tilde{U}$ and $\Tilde{V}$ with the ground truth data $U_\mathrm{m}$ and $V_\mathrm{m}$ at cross-sections corresponding to all gauge positions $x_\mathrm{g} \in \{3, 4, 5, 6\} \, \mathrm{m}$.  The average across all four gauges yields a value of $\mathrm{SSP}=0.0728$ for the real part and $\mathrm{SSP}=0.1071$ for the imaginary part. 
In the lower subplots, we observe precise adherence to the boundary conditions at $x_\mathrm{g,maes} \in \{3, \, 6\} \, \mathrm{m}$, while the reconstruction within the domain at $x_\mathrm{g, test} \in \{4, \, 5 \}$ exhibits slightly less agreement with the measurements, yet consistently maintains low errors of $\mathrm{SSP}\le 0.15$. This aligns with the observed plateau of PDE errors in the loss curve in Fig.~\ref{fig:example_loss}, given that the PINN solution within the computational domain is primarily driven by the PDE-residual loss components. Concerning the simplified assumptions inherent in the NLSE, it is crucial to note that the ground truth generated by the more realistic HOSM may only be approximated to a certain extent due to model misspecification.

\begin{figure}[ht!]
     \centering
     \includegraphics[width=0.97\textwidth]{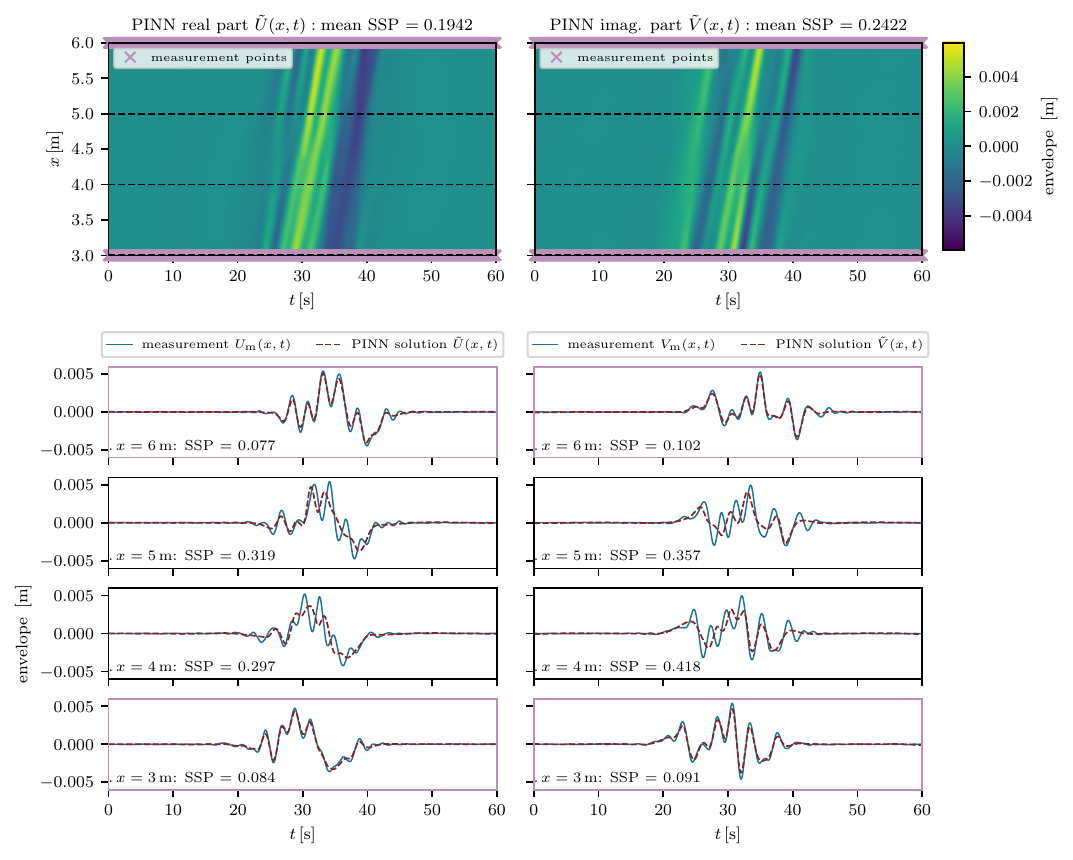}
      \caption{Real part (a) and imaginary part (b) of the envelope solution $\Tilde{A}(x,t) = \Tilde{U}(x,t) + i \Tilde{V}(x,t)$ generated by the PINN using predetermined, constant NLSE coefficients for sample no. 2 (carrier wave with $\epsilon=0.025$, $\overline{\omega}_\mathrm{p}= 8.34 \, \frac{\mathrm{rad}}{\mathrm{s}}$, $\gamma=1$). Compared to sample no. 1 in Fig. \ref{fig:samp1}, the envelope maxima developing from the boundaries meet with an offset in the middle of the domain.}
     \label{fig:samp2}
\end{figure}

While notable success is achieved in the reconstruction of certain wave samples, challenges are evident for other samples. For instance, Fig.~\ref{fig:samp2} presents a PINN solution for sample no. 2 described by $\epsilon=0.025$, $\overline{\omega}_\mathrm{p}=8.34 \frac{\mathrm{rad}}{\mathrm{s}}$, which is broad-banded as $\gamma=1$. The average errors for the real and imaginary parts of the envelope are $\mathrm{SSP}=0.1942$ and $\mathrm{SSP}=0.2422$, respectively. Remarkable errors occur within the computational domain at $x_\mathrm{g,test} \in \{ 4,\, 5 \} \, \mathrm{m}$, despite satisfactory alignment of boundary values with the ground truth at  $x_\mathrm{g, meas} \in \{3, \, 6 \} \, \mathrm{m}$. We make the nonphysical observation of dominant envelope maxima developing from both boundaries towards the middle of the domain, but meeting with an offset. This offset indicates a small error in the calculated group velocity of the NLSE $c_\mathrm{g} = \frac{\overline{\omega}_\mathrm{p}}{2 \overline{k}_\mathrm{p}}$ for the underlying data. This discrepancy may arise from challenges in determining the peak frequency $\overline{\omega}_\mathrm{p}$ from wave spectra as wave information in segments of less than 60 seconds is provided.

\begin{figure}[ht!]
\centering
\includegraphics{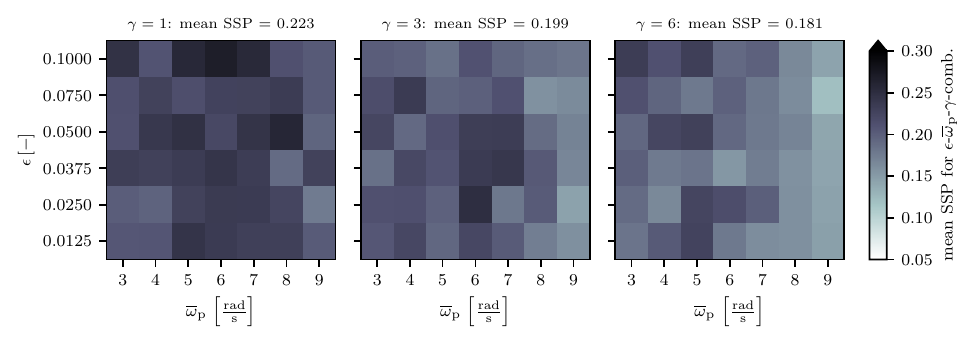}
\vspace{-0.5cm}
     \caption{Reconstruction errors for all 630 samples from training the PINNs with predetermined, constant NLSE coefficients. In general, the envelopes above lower-frequency waves (lower $\overline{\omega}_\mathrm{p}$) are harder to reconstruct than for higher-frequency waves. Moreover, the errors increase for broader-banded samples (lower $\gamma$) and higher steepness samples (higher $\epsilon$). }
     \label{fig:error_surfaces}

\end{figure}

The trends visually presented in the two examples above are repetitive across all 630 instances in the assimilation task. Results for the remaining samples are summarized in Fig.~\ref{fig:error_surfaces}, where the SSP value represents the average error for the real and imaginary part of each sample. Moreover, each cell represents the mean SSP achieved for all five samples of a $\epsilon$-$\overline{\omega}_\mathrm{p}$-$\gamma$-combination, whereas the determined $\overline{\omega}_\mathrm{p}$ value is rounded to the nearest integer. Notably, errors tend to decrease as the $\overline{\omega}_\mathrm{p}$-value increases, which may seem counterintuitive given the PINN method's successive solving from lower to higher frequency components \citep[cf.][]{Cuomo2022}. However, this is explainable by the observation that higher carrier wave frequencies $\overline{\omega}_\mathrm{p}$ result in smoother, lower-frequency envelopes, while lower-frequency carrier waves yield less smooth envelopes. As we are reconstructing wave envelopes and not the wave elevation itself, the errors' dependence on the $\overline{\omega}_\mathrm{p}$-value aligns with the frequency dependence of the PINN method.
Moreover, Fig.~\ref{fig:error_surfaces} illustrates a general increase in errors as the peak enhancement factors $\gamma$ decrease or the steepness values $\epsilon$ slightly increase. These observations are not attributable to the PINN method itself but align with the model misspecification: The NLSE is limited for narrow-band and small-amplitude waves, while the ground truth HOSM data seems to exceed the validity range of the NLSE for increasing $\epsilon$ and broader spectra such as samples with $\gamma = 1$. Accordingly, our investigations validate Hypothesis \ref{hyp:1}, affirming that gauge reduction through data assimilation is feasible, as long as the measurement data closely adheres to the limitations of the NLSE. In our case, this is achievable for the medium- and narrow-banded spectra of $\gamma=3$ and $\gamma=6$, yielding average error values of $\mathrm{SSP}=0.199$ and $\mathrm{SSP}=0.181$. However, the approach encounters limitations for broad-banded wave fields defined by $\gamma=1$ resulting in $\mathrm{SSP}=0.223$, surpassing the defined threshold of $\mathrm{SSP} \le 0.2$ for satisfactory reconstruction quality. Additionally, it is crucial to note that reconstructing wave samples with even higher steepness values than the considered $\epsilon$ will further increase errors.

\subsection{Coefficient identification}
\label{subsec:coefficient identification}

Despite the advanced method to determine the peak frequency $\overline{\omega}_\mathrm{p}$ with Eq.~\eqref{eq:omega_ref}, which is crucial for calculating NLSE coefficients, uncertainties in the $\overline{\omega}_\mathrm{p}$-value persist, as discussed in the context of the envelope offset observed in Fig.~\ref{fig:samp2}. Due to the impracticality of exploring alternative possibilities for $\overline{\omega}_\mathrm{p}$-determination or iteratively adjusting it to recalculate the solution, we examine a data-driven opportunity in the following subsection.  The PINN methodology allows treating PDE coefficients as additional trainable variables, addressing inverse problems during assimilation of measurement data  \citep[cf.][]{Raissi2017}. We aim to enhance the reconstruction quality (Hypothesis~\ref{hyp:2}) and eliminate envelope offsets attributed to non-optimal coefficients. As in the NLSE's case, all coefficients are linked to $\overline{\omega}_\mathrm{p}$ and $\overline{k}_\mathrm{p}$, we refrain from directly setting $c_\mathrm{g}$, $\delta$ and $\nu$ as variables. Instead, we initially determine $\overline{\omega}_\mathrm{p}$ and $\overline{k}_\mathrm{p}$ using Eqs. \eqref{eq:omega_ref} and \eqref{eq:linear_dispersion} as done before but subsequently treat them as additional PINN variables.
Moreover, as water waves with increasing nonlinearity are known to deviate from the linear dispersion relation (Eq. \eqref{eq:linear_dispersion}), we decouple the current $\overline{k}_\mathrm{p}$ and $\overline{\omega}_\mathrm{p}$ during training. 
As done in the previous subsection, each of the 630 samples undergoes an individual training for 15,000 epochs, now incorporating two additional trainable variables $\overline{\omega}_\mathrm{p}$ and $\overline{k}_\mathrm{p}$. \\

\begin{figure}[ht!]
     \centering
     \includegraphics[width=0.97\textwidth]{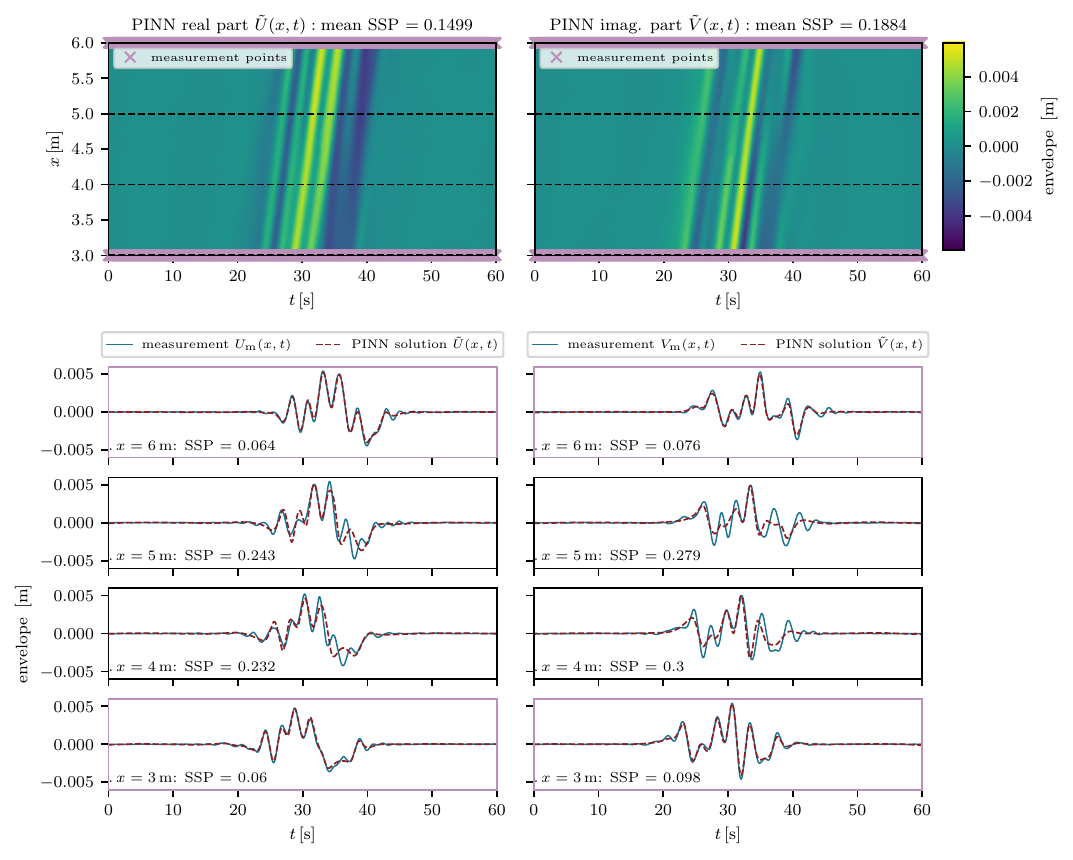}
         \caption{Real part (a) and imaginary part (b) of the envelope solution $\Tilde{A}(x,t) = \Tilde{U}(x,t) + i \Tilde{V}(x,t)$ generated by the PINN using adaptable NLSE coefficients due to learnable $\overline{\omega}_\mathrm{p}$ and $\overline{k}_\mathrm{p}$ for sample no. 2 (carrier wave with $\epsilon=0.025$, $\overline{\omega}_\mathrm{p}= 8.34 \, \frac{\mathrm{rad}}{\mathrm{s}}$, $\gamma=1$). Compared to the reconstruction using constant NLSE coefficients in Fig. \ref{fig:samp2}, a reduction in the envelope offset is evident.}
     \label{fig:samp2_omega_free}
\end{figure}

Comparing the new reconstruction for sample no. 2 employing trainable variables $\overline{\omega}_\mathrm{p}$ and  $\overline{k}_\mathrm{p}$ in Fig.~\ref{fig:samp2_omega_free} with the previous solution using constant coefficients in Fig.~\ref{fig:samp2}, reveals a remarkable reduction in the envelope offset in the top view and an improved alignment of phases and amplitudes with the ground truth. Table~\ref{tab:omega_identify} presents a comparison between the initial peak frequency and wavenumber extracted from the spectrum and the values learned during training, including the corresponding NLSE coefficients. For this sample, the $\overline{\omega}_\mathrm{p}$ variable increased during training while the $\overline{k}_\mathrm{p}$ decreased, resulting in a faster group velocity $\scriptstyle{c_\mathrm{g}= \tfrac{\overline{\omega}_\mathrm{p}}{2 \overline{k}_\mathrm{p}}}$. Simultaneously, the nonlinear term in the NLSE receives slightly less attention due to decreased coefficient $\scriptstyle{\nu=- \overline{k}_\mathrm{p}^3}$. These adjustments result in improved error values for the real part, decreasing from $\mathrm{SSP}=0.1942$ to $\mathrm{SSP}=0.1499$ and for the imaginary part, decreasing from $\mathrm{SSP}=0.2422$ to $\mathrm{SSP}=0.1884$. This improvement is acceptable, considering that adjustable coefficients can mitigate uncertainties in $\overline{\omega}_\mathrm{p}$-determination but cannot overcome the NLSE's assumption of a single carrier frequency, which is not valid, especially for HOSM data with $\gamma=1$.

\begin{table}[!ht]
\centering
\caption{NLSE coefficients calculated from $\overline{\omega}_\mathrm{p}$ and $\overline{k}_\mathrm{p}$ at the beginning of the PINN training compared to the optimized values learned at the end. This table displays the values for sample no. 2, whose reconstruction result is shown in Fig. \ref{fig:samp2_omega_free}.}
\begin{footnotesize}
\begin{tabular}{lccccccc}
\toprule
\toprule
\multicolumn{3}{c}{} & \multicolumn{3}{c}{NLSE coefficients}   \\
\cmidrule(rl){4-6} 
 & $\overline{\omega}_\mathrm{p}$ & $\overline{k}_\mathrm{p}$ & $c_\mathrm{g}=\tfrac{\overline{\omega}_\mathrm{p}}{2 \overline{k}_\mathrm{p}}$ & $\delta= - \tfrac{\overline{k}_\mathrm{p}}{\overline{\omega}^2_\mathrm{p}}$ & $\nu= - \overline{k}_\mathrm{p}^3$ \\ 
\midrule
\midrule
initial  & 8.344   & 7.097  & 0.588 & -0.102    & -357.4 \\
learned  & 9.355   & 6.202  & 0.754 & -0.071    & -256.7  \\
\bottomrule
\bottomrule
\end{tabular}
\end{footnotesize}
\label{tab:omega_identify}
\end{table}

\begin{figure}[ht!]
\centering
\includegraphics{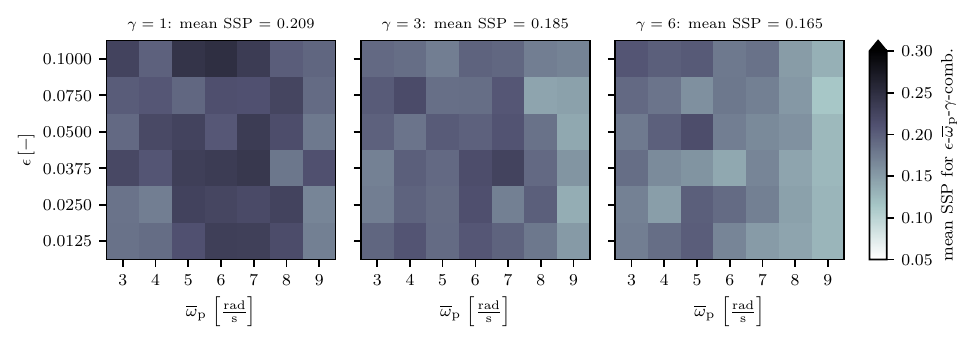}
\vspace{-0.5cm}
     \caption{Reconstruction errors for all 630 samples from training the PINNs with learnable $\overline{\omega}_\mathrm{p}$- and $\overline{k}_\mathrm{p}$-values causing adaptable NLSE coefficients. In general, the SSP errors are slightly reduced compared to results obtained with constant coefficients in Fig.~\ref{fig:error_surfaces}. However, the general trend, that lower $\overline{\omega}_\mathrm{p}$-values, higher $\epsilon$-values and lower $\gamma$-values lead to increased errors remains.}
     \label{fig:error_surfaces_omega_free}
\end{figure}

Examining the 630 samples collectively reveals that the solutions using trainable $\overline{\omega}_\mathrm{p}$ and $\overline{k}_\mathrm{p}$ values exhibit fewer envelope offsets. Fig. \ref{fig:error_surfaces_omega_free} demonstrates a notable enhancement in SSP values when comparing these results for adaptable coefficients to results obtained with constant and predetermined coefficients shown in Fig. \ref{fig:error_surfaces}.  While the trend persists, where lower-frequency, higher-steepness, or broad-banded samples yield comparatively high individual errors, the average SSP values experience a reduction from $\mathrm{SSP}=0.223$ to $\mathrm{SSP}=0.209$ ($\gamma=1$), from $\mathrm{SSP}=0.199$ to $\mathrm{SSP}= 0.185$ ($\gamma=3$) and from $\mathrm{SSP}=0.181$ to $\mathrm{SSP}=0.165$ ($\gamma=6$). On average the reduction from from $\mathrm{SSP}=\frac{1}{3}(0.223+0.199+0.181)=0.204$ with constant coefficients to $\mathrm{SSP}=\frac{1}{3}(0.209+0.185+0.165)=0.186$ with trainable coefficients represents an improvement of approximately $8.82\%$.  Consequently, the utilization of trainable NLSE coefficients confirms Hypothesis~\ref{hyp:2}, as it enhances measurable and visible reconstruction quality compared to constant coefficients.

\begin{figure}[ht!]
\vspace{-0.3cm}
\centering
\includegraphics[width=0.76\textwidth]{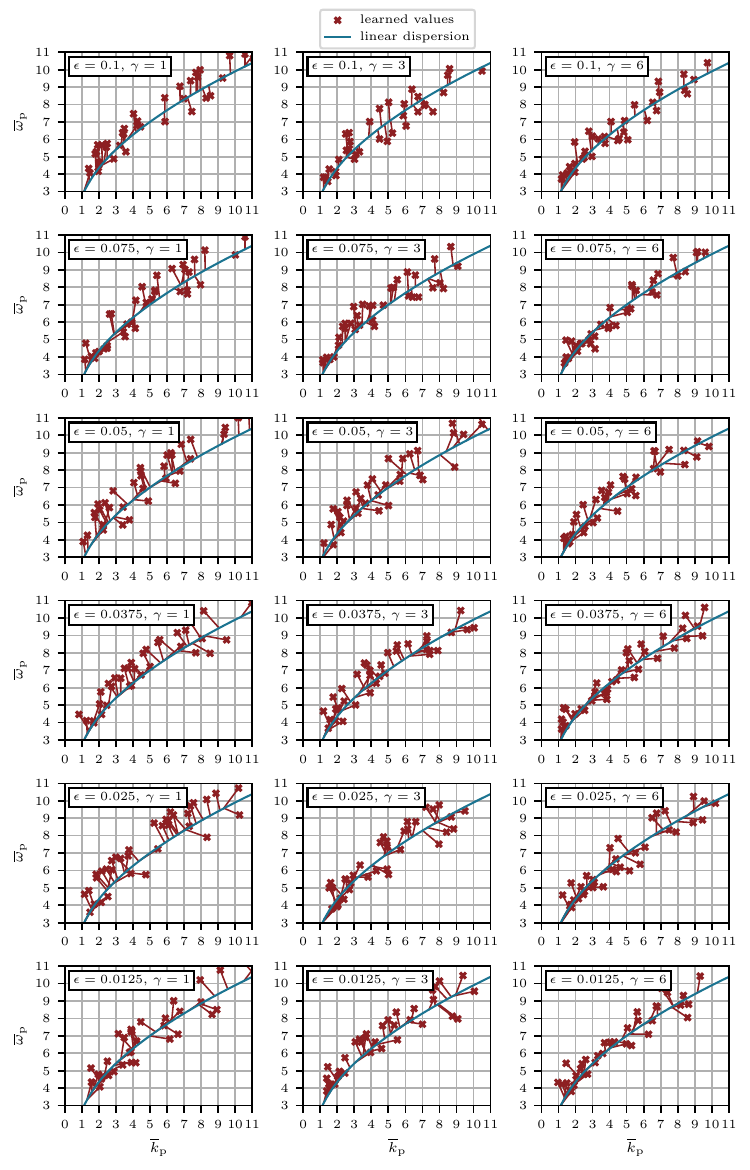}
\vspace{-0.2cm}
     \caption{Initial and final $\overline{\omega}_\mathrm{p}$-$\overline{k}_\mathrm{p}$-combinations for all 630 samples from training the PINNs with learnable $\overline{\omega}_\mathrm{p}$ and $\overline{k}_\mathrm{p}$ values causing adaptable NLSE coefficients. While the initial values are determined following the linear dispersion relation, this constraint is removed during training. We observe hat braoder-banded samples (lower $\gamma$) tend to benefit from $\overline{\omega}_\mathrm{p}$-$\overline{k}_\mathrm{p}$-combinations above linear dispersion. Furthermore, with increasing steepness $\epsilon$, the PINNs attempt to remain the nonlinear term of the NLSE associated with $\scriptstyle{\nu=- \overline{k}_\mathrm{p}^3}$ as far as possible by allowing minor changes to the $\overline{k}_\mathrm{p}$ value.   }
     \label{fig:nonlinear_dispersion}
\end{figure}

During the coefficient identification task, the adaption of peak frequency and wavenumber is not universally consistent among the 630 samples. For the sample shown in Fig. \ref{fig:samp2_omega_free} and Tab. \ref{tab:omega_identify}, the peak frequency increased and the wavenumber decreased during training. Other samples show variations, such as increasing both $\overline{\omega}_\mathrm{p}$ and $\overline{k}_\mathrm{p}$, decreasing both $\overline{\omega}_\mathrm{p}$ and $\overline{k}_\mathrm{p}$ or decreasing $\overline{\omega}_\mathrm{p}$ while increasing $\overline{k}_\mathrm{p}$. Fig \ref{fig:nonlinear_dispersion} summarizes this diversity for all $\epsilon$-$\gamma$-combinations. Initial $\overline{\omega}_\mathrm{p}$ and $\overline{k}_\mathrm{p}$ values are determined according to the blue graph of the linear dispersion relation, while the learned values mostly deviate from this relation, as indicated by red lines and crosses. The left and middle column of this figure reveal that medium- and broader-banded sea states ($\gamma=1$ and $\gamma=3$) tend to benefit from $\overline{\omega}_\mathrm{p}$-$\overline{k}_\mathrm{p}$-combinations above the linear dispersion relation. This suggests that slightly higher group velocities $\scriptstyle{c_\mathrm{g}= \tfrac{\overline{\omega}_\mathrm{p}}{2 \overline{k}_\mathrm{p}}}$ enhance reconstruction quality compared to values obtained from Eq. \eqref{eq:omega_ref} and linear dispersion. In contrast, for narrow-banded samples in the right column ($\gamma=6$), the learned $\overline{\omega}_\mathrm{p}$-$\overline{k}_\mathrm{p}$-combinations occasionally fall above or below the linear dispersion curve, reflecting reduced uncertainties in the initial determination of $\overline{\omega}_\mathrm{p}$.\\

Moreover, Fig. \ref{fig:nonlinear_dispersion} reveals the trend, that independent of the $\gamma$-value, increasing steepness $\epsilon$ leads to smaller changes in the $\overline{k}_\mathbf{p}$ values during training. This is evident from the vertically oriented lines between initial and final $\overline{\omega}_\mathrm{p}$-$\overline{k}_\mathrm{p}$-combinations in the upper subplots. In contrast, in the lower subplots for samples with smaller steepness $\epsilon$, changes in both smaller and larger $\overline{k}_\mathrm{p}$ values are observed. This implies that as steepness increases, the PINN inherently prioritizes maintaining the nonlinear term of the NLSE, represented by the coefficient $\scriptstyle{\nu=-\overline{k}^3_\mathrm{p}}$. The observation aligns with physical reasoning, as higher $\epsilon$ correspond to increased nonlinearities through higher amplitudes. Preserving this coefficient thus indicates that the nonlinearity is identified as crucial by the PINN. Consequently, alterations in group velocity $\scriptstyle{c_\mathrm{g}=\tfrac{\overline{\omega}_\mathrm{p}}{2 \overline{k}_\mathrm{p}}}$ primarily occur through alterations in $\overline{\omega}_\mathrm{p}$ only. 

\section{Conclusion}
\label{sec:conclusion}

In this study, we showcase practical applications of physics-informed neural networks (PINNs) for questions arising in ocean engineering and water wave research. The hydrodynamic nonlinear Schrödinger equation constrains the developed PINN framework to ensure the physical consistency of its solutions. Our first investigation demonstrates the PINN's ability to reduce the number of costly gauges in wave tank experiments by reconstructing wave elevation envelopes between two measurement points, a process known as data assimilation. However, uncertainties in the a-priori determination of constant NLSE coefficients in this task occasionally lead to non-physical envelope reconstructions, as indicated by offsets in certain instances.

To enhance the reconstruction quality of such instances, conventional numerical methods would typically require a systematic coefficient adjustment and numerical forward propagation to optimize the data fit between two measurement points. In contrast, our NLSE-PINN method allows the direct identification of optimized NLSE coefficients to align with provided boundary data, as demonstrated in the second phase of our research. Considering that all NLSE coefficients are related to the samples' peak frequency and peak wavenumber, we select $\overline{\omega}_\mathrm{p}$ and $\overline{k}_\mathrm{p}$ as trainable variables. This additional coefficient identification task allows us to reduce the mean error across all 630 wave samples from $\mathrm{SSP}=0.204$ with predetermined, constant coefficients to $\mathrm{SSP}=0.186$ with trainable coefficients, representing an improvement of $8.82$ percent. Moreover, incorporating trainable coefficients results in a remarkable reduction of non-physical envelope offsets observed before. However, it is important to note that the reconstruction remains generally more challenging for broader wave spectra and higher steepness samples. This challenge is not inherent to the PINN methodology but is associated with the characteristics of the utilized NLSE to constrain the loss function: The classical hydrodynamic NLSE is defined for narrow-banded wave packets with moderate steepness. In contrast, the synthetic wave measurement data is generated using the high-order spectral method valid for a wider range of sea states.

For this reason, an avenue for future research is the development of a PINN capable of directly solving the fully nonlinear potential flow equations instead of relying on simplified wave equations such as the NLSE. However, this objective requires addressing computational challenges associated with simultaneously approximating the potential field $\Phi(x,z,t)$ and wave elevation $\eta(x,t)$. This involves managing a substantial number of collocation points not only across the sea surface but also in the depth direction.

In addition, concerning the NLSE coefficient identification task in this study, a promising avenue for further research is to explore the feasibility of deriving a nonlinear dispersion relation as a function of wave parameters, such as bandwidth, amplitude, or steepness. Subsequent analysis could assess whether this identified NLSE system can mitigate restrictions in bandwidth and steepness by computing solutions for $x>6 \, \mathrm{m}$ and comparing its performance with solutions derived from modified NLSE variants \citep[cf.][]{Dysthe1979, Trulsen1996}. This research direction has the potential to deepen our understanding of the interplay of NLSE coefficients and wave parameters and may contribute to the development of more flexible and accurate models for describing nonlinear water wave phenomena.

Furthermore, it is crucial to note that this study employed synthetic wave data generated within a numerical wave tank. Hence, future research efforts could explore the application of the potential flow PINN or NLSE-PINN method to real wave measurement data obtained from wave tanks or even in open ocean scenarios, particularly when considering directional two-dimensional wave surfaces.


\newpage
\section*{Funding}
This work was supported by the Deutsche Forschungsgemeinschaft (DFG - German Research Foundation) [project number 277972093: Excitability of Ocean Rogue Waves]

\section*{Acknowledgements}
The authors would like to express their gratitude to the lecturers George Em. Karniadakis and Khemraj Shukla, as well as to all organizers and funders for the Summer School on Physics-Informed Neural Networks and Applications at KTH, Stockholm in 2023. Furthermore, appreciation is extended to the Studienstiftung des deutschen Volkes for the support provided through the Natural Science Collegium 2022/2023, during which time this work was initiated.

\section*{Declaration of interests}

The authors declare that they have no known competing financial interests or personal relationships that could have appeared to influence the work reported in this paper.

\section*{Declaration of g enerative AI and AI-
assisted technologies in the writing process}

The investigations were conducted and the manuscript was written completely by the authors. Once the manuscript was completed, the authors used ChatGPT 3.5 (\url{https://chat.openai.com/}) in order to improve its grammar and readability. After using this tool, the authors reviewed and edited the content as needed and take full responsibility for the content of the publication.

\bibliography{bibliography}

\end{document}